\documentclass{article}

\usepackage{PRIMEarxiv}
\usepackage{amsmath,graphicx}
\usepackage{cite}
\usepackage{amsmath,amssymb,amsfonts}
\usepackage{algorithmic}
\usepackage{textcomp}
\usepackage{xcolor}
\usepackage{booktabs}
\usepackage{url}
\usepackage{amssymb}

\newcommand{\tabref}[1]{{Table \ref{#1}}}

\newcommand{\secref}[1]{{Section \ref{#1}}}

\newcommand{\figref}[1]{{Figure \ref{#1}}}

\title{Towards Audio Domain Adaptation for Acoustic Scene Classification using Disentanglement Learning\thanks{Submitted to ICASSP 2021}
}

\author{
  Jakob Abe{\ss}er \\
  Semantic Music Technologies Group \\
  Fraunhofer IDMT \\
  Ilmenau, Germany\\
  \texttt{jakob.abesser@idmt.fraunhofer.de} \\
   \And
  Meinard M{\"u}ller \\
  International Audio Laboratories \\
  Erlangen, Germany\\
  \texttt{meinard.mueller@audiolabs-erlangen.de} \\
}

\begin{document}
\maketitle
\begin{abstract}
The deployment of machine listening algorithms in real-life applications is often impeded by a domain shift caused for instance by different microphone characteristics.
In this paper, we propose a novel domain adaptation strategy based on disentanglement learning. 
The goal is to disentangle task-specific and domain-specific characteristics in the analyzed audio recordings.
In particular, we combine two strategies: First, we apply different binary masks to internal embedding representations and, second, we suggest a novel combination of categorical cross-entropy and variance-based losses.
Our results confirm the disentanglement of both tasks on an embedding level but show only minor improvement in the acoustic scene classification performance, when training data from both domains can be used.
As a second finding, we can confirm the effectiveness of a state-of-the-art unsupervised domain adaptation strategy, which performs across-domain adaptation on a feature-level instead.
\end{abstract}

%
\section{Introduction}

Acoustic scene classification (ASC) is an essential task of auditory scene analysis. Its goal is to classify the audio recording's environment according to a given set of pre-defined classes. Examples of such classes are indoor locations (e.\,g., restaurants and metro stations) and outdoor locations (e.\,g., pedestrian streets and parks).
State-of-the-art ASC algorithms are mostly driven by deep learning (DL) techniques and, to this day, face several challenges in real-world application scenarios, see \cite{Abesser:2020:ASC:AS} for a summary.

A key challenge is that the data used to train ASC algorithms is typically recorded under different acoustic conditions (microphone characteristics, reverberations, background noises) than the audio data recorded by the potential application device.
Such a microphone mismatch can cause a ``domain shift,'' which is a distribution mismatch between the source domain (training) data and the target domain (test) data to be expected in the application scenario.
It has been shown in related disciplines like computer vision that such domain shifts can cause significant performance degradation of (deep) neural networks \cite{Kouw:2018:DomainAdaptation:ARXIV, Wang:2018:DomainAdaptation:NC}.
Several domain adaptation (DA) strategies have been proposed in the past as countermeasures. Amongst others, these strategies include a fine-tuning of pre-trained classification models onto target domain data (transfer learning) or an adaptation/normalization of the input features without changing the initial model.
From a practical viewpoint, both strategies require an additional adaptation effort every time an ASC system faces audio data from a novel target domain.

As the main contribution of this paper, we propose to use disentanglement representation learning during the model training process to learn task-specific yet domain-agnostic feature representations.
Our assumption is that there exist two main  factors which influence the acoustic variability of multi-device acoustic scene recordings. First, there are domain-specific properties such as microphone characteristics and room acoustics, which do not carry meaningful information but rather confuse ASC algorithms. The second factor are the relevant characteristics of different acoustic scenes.
By learning domain-agnostic feature representations, our objective is to develop more robust and context-sensitive ASC approaches, which can be used for devices such as hearing aids and cellphones to adapt to changes in the surrounding acoustic environment automatically.
In order to foster scientific reproducibility, we publish Python code to reproduce our experiments.\footnote{\url{https://github.com/jakobabesser/ICASSP_2022_DisentanglementASC}}

\section{Related Work}

Modern ASC approaches almost exclusively rely on  use convolutional neural networks (CNN) or convolutional recurrent neural networks (CRNN) to learn discriminative features in time-frequency audio representations such as Mel spectrograms. We refer the reader to  \cite{Abesser:2020:ASC:AS} for a survey on state-of-the-art ASC approaches.
Several publications proposed DA methods for audio signals to compensate for domain shift caused, for instance, by microphone mismatch conditions.

DA methods can be categorized into unsupervised and supervised methods depending on whether labels of the target domain data exist or not.
In supervised DA scenarios, ASC algorithms can be trained jointly for ASC and domain classification (DC) in a multitask learning fashion \cite{Denisov:2020:DA:SC, Tang:2021:SED:ARXIV}.
A similar multitask learning approach can also be used for semi-supervised learning, as shown in \cite{Lee:2021:ContrastiveRegularizationDA:WASPAA}. 

For unsupervised DA, adversarial training strategies were proposed to align the distributions of intermediate feature representations in ASC models obtained from source and target domain data \cite{Gharib:2018:DomainAdaptationASC:DCASE, Drossos:2019:DomainAdaptation:WASPAA}.
or by mapping data from both domains to another (universal) domain \cite{Mun:2019:DomainAdaptation:ICASSP, Mezza:2020:DomainAdaptation:MLSP}. 
One approach to achieve such an alignment is to standardize data per frequency band independently per domain \cite{Johnson:2020:Normalization:EUSIPCO} or by matching the band-wise statistics between domains \cite{Mezza:2021:DomainAdaptation:EUSIPCO}.
In general, such an adaptation can not only be performed on the feature-level but also using  internal hidden layer activations \cite{Primus:2019:DeviceInvariantASC:DCASE}.

In order to compensate for differences in the microphone frequency responses, Kosmider~\cite{Kosmider:2019:DeviceCalibration:DCASE} propose a ``spectral correction'' step. Based on time-aligned recordings from different recording devices (each one is considered as one domain), frequency-dependent magnitude coefficients are estimated from the source data and used to calibrate the target domain data.
Mezza et al.~\cite{Mezza:2020:DomainAdaptation:MLSP} proposed to project source and target domain features to a lower-dimensional subspace spanned by the eigenvectors of the source domain feature covariance matrix. 
Hu et al.~ \cite{Hu:2020:DeviceAdaptationASC:INTERSPEECH}  propose to use neural label embedding (NLE) to encode structural relationships between different acoustic scene classes from source domain data. In order to mitigate microphone mismatch, this knowledge is then transferred to target domain data using relational teacher--student learning.

A promising approach to train more robust classification models is disentanglement representation learning.  
Mun and Shon~\cite{Mun:2019:DomainAdaptation:ICASSP} proposed to use a factorized hierarchical variational autoencoder (FHVAE) in order to first disentangle the audio features' component related to the recording device, shift it according to a universal domain mean value, and finally decode and reconstruct the modified audio features.
In the field of music information retrieval, 
Lee et al.~\cite{Lee:2020:Disentanglement:ISMIR} used a Conditional Similarity Network (CSN) \cite{Veit:2017:CSN:ARXIV} to disentangle multiple semantic concepts such as genre, mood, or instrumentation in music recordings in order to learn configurable similarity metrics. The authors applied a binary masking procedure on internal embedding representations in order to enforce a specific distribution of semantic concepts within this embedding. We adapt this approach to disentangle the ASC and DC components as will be detailed in the next section.
As the main conceptual difference to previous DA approaches, we aim to learn domain-agnostic feature representations already at the training stage, a strategy that does not require any further adaptation when novel target domains are faced. 

\section{Proposed Method}

In this section, we first introduce in \secref{sec:dataset} the ASC dataset used for experimental validation. Then, we describe in \secref{sec:emb_norm_and_masking} the applied neural network architecture, which implements a multitask learning approach of ASC and DC as well as an embedding masking procedure in order to disentangle both concepts.
Finally, \secref{sec:training_loss_configurations} summarizes different training configurations, covering regular ASC training as well as unsupervised and supervised DA scenarios.

\subsection{Dataset \& Feature Extraction}
\label{sec:dataset}

In this paper, we focus on the DCASE 2018 Task 1B entitled ``Acoustic Scene Classification with mismatched recording devices'', which addresses ASC under microphone mismatch conditions. 
The corresponding dataset includes ten second long excerpts of acoustic scene recordings (denoted as ``clips'' in the following) recorded with three devices: a source domain device (A) as well as two target domain devices (B \& C).
For the sake of reproducibility, we use in our experiments a set of pre-computed Mel spectrogram features extracted from the task's development set, which were published alongside with \cite{Gharib:2018:DomainAdaptationASC:DCASE}\footnote{\url{https://zenodo.org/record/1401995}}. 
These features cover $F=64$ Mel bands and were computed using a window size of 2048 samples with a 50~\% overlap.
This feature set was used in follow-up publications on unsupervised DA \cite{Drossos:2019:DomainAdaptation:WASPAA, Mezza:2021:DomainAdaptation:EUSIPCO}.
Based on the provided dataset split, we use the training set with 5510/612/612 clips for devices A/B/C for model training and the test set with 2518/180/180 clips for evaluation.
Throughout this paper, the subscript ``$\mathrm{A}$'' indicates the acoustic scene and ``$\mathrm{D}$'' the domain classification tasks. Furthermore, the superscripts ``$\mathrm{S}$'' and ``$\mathrm{T}$'' refer to the source and target domain, respectively. We consider $A=10$ acoustic scene classes and $D=3$ domain classes.

\subsection{Embedding Normalization \& Masking}
\label{sec:emb_norm_and_masking}

\begin{figure}[t] 
	\begin{center}
		\includegraphics[width=0.5\textwidth]{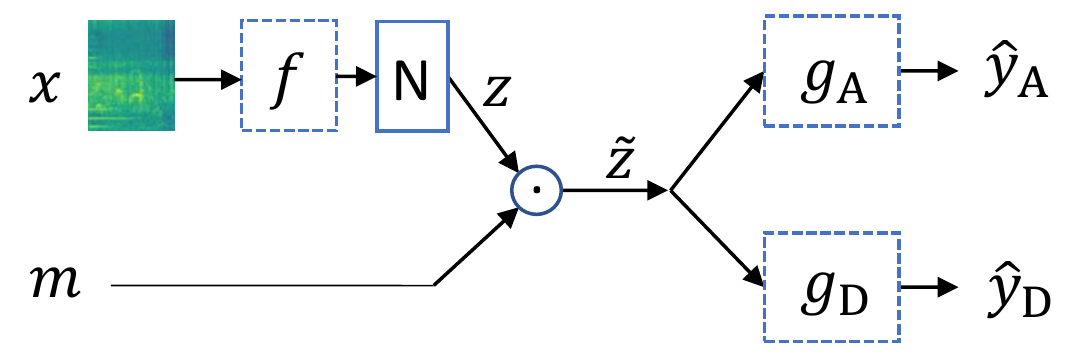}
		\caption{General neural network architecture: A core model $f$ transforms input features $x$ to a normalized (N) embedding representation $z$, which is multiplied (element-wise) with a binary mask $m$. The masked embedding $\tilde{z}$ is processed by two output branches to obtain acoustic scene and domain predictions $\hat{y}_\mathrm{A}$ and $\hat{y}_\mathrm{D}$, respectively. Trainable and fixed parts of the network are shown with dashed and solid blue rectangles, respectively.}
		\label{fig:flowchart}
	\end{center}
\end{figure}
\figref{fig:flowchart} gives an overview our network architecture. Each clip is represented by a feature tensor $x~\in~\mathbb{R}^{T \times F \times C}$ covering $T=431$ time frames, $F=64$ Mel bands, and one depth channel ($C=1$).
The output of the core model $f: \mathbb{R}^{T \times F \times C} \to  \mathbb{R}^E$ is an embedding vector $z = f(x)$.
Zhai et al.~\cite{Zhai:2020:ClassificationMetricLearning:BMVC} discussed the importance of normalizing embedding vectors when being used for classification tasks. Therefore, we apply layer normalization \cite{Ba:2016:LayerNorm:ARXIV} to normalize each embedding vector independently to zero mean and unit variance. 
After the normalization step, each embedding $z$ is multiplied element-wise with a binary mask $m \in \left\{0,1\right\}^{E}$ yielding $\tilde{z} = z \odot m$.
Since we aim to disentangle both semantic concepts (acoustic scene and domain), we organize $m$ such that its first half contains zeros and the second half contains ones 
or vice versa. The main intuition is that only the unmasked parts of the embedding vector can provide information for each of the two classification tasks. As a consequence, this task specialization should support the disentanglement \cite{Lee:2020:Disentanglement:ISMIR}.


The masked embedding vectors $\tilde{z}$ are propagated to two output branches each with one dense layer with softmax activation in order to predict the acoustic scene and the audio domain. The output branch operations are denoted as $g_\mathrm{A}$ and $g_\mathrm{D}$ and the acoustic scene and domain predictions $\hat{y}_\mathrm{A}~\in~\mathbb{R}^{A}$ and $\hat{y}_\mathrm{D}~\in~\mathbb{R}^{D}$ are computed as $\hat{y}_\mathrm{A}=g_\mathrm{A}\left(\tilde{z}\right)$ and $\hat{y}_\mathrm{D}=g_\mathrm{D}\left(\tilde{z}\right)$, respectively. 
Therefore, the model training involves learning $f$, $g_\mathrm{A}$, and $g_\mathrm{D}$ from batches of data instances, which consist of feature-mask pairs $(x, m)$ and the corresponding targets $(y_\mathrm{A}, y_\mathrm{D})$.

For the sake of comparability with prior research, we use the ``Kaggle'' CNN model, which is detailed in Table II in \cite{Mezza:2021:DomainAdaptation:EUSIPCO}.
The network includes five convolutional layers followed by two dense layers each with 256 units.
We use this model as core model $f$ and consider the output of the second dense layer as embedding $z$.

\subsection{Training Strategies \& Loss Configurations}
\label{sec:training_loss_configurations}

\begin{figure}[t] 
	\begin{center}
		\includegraphics[width=0.7\textwidth]{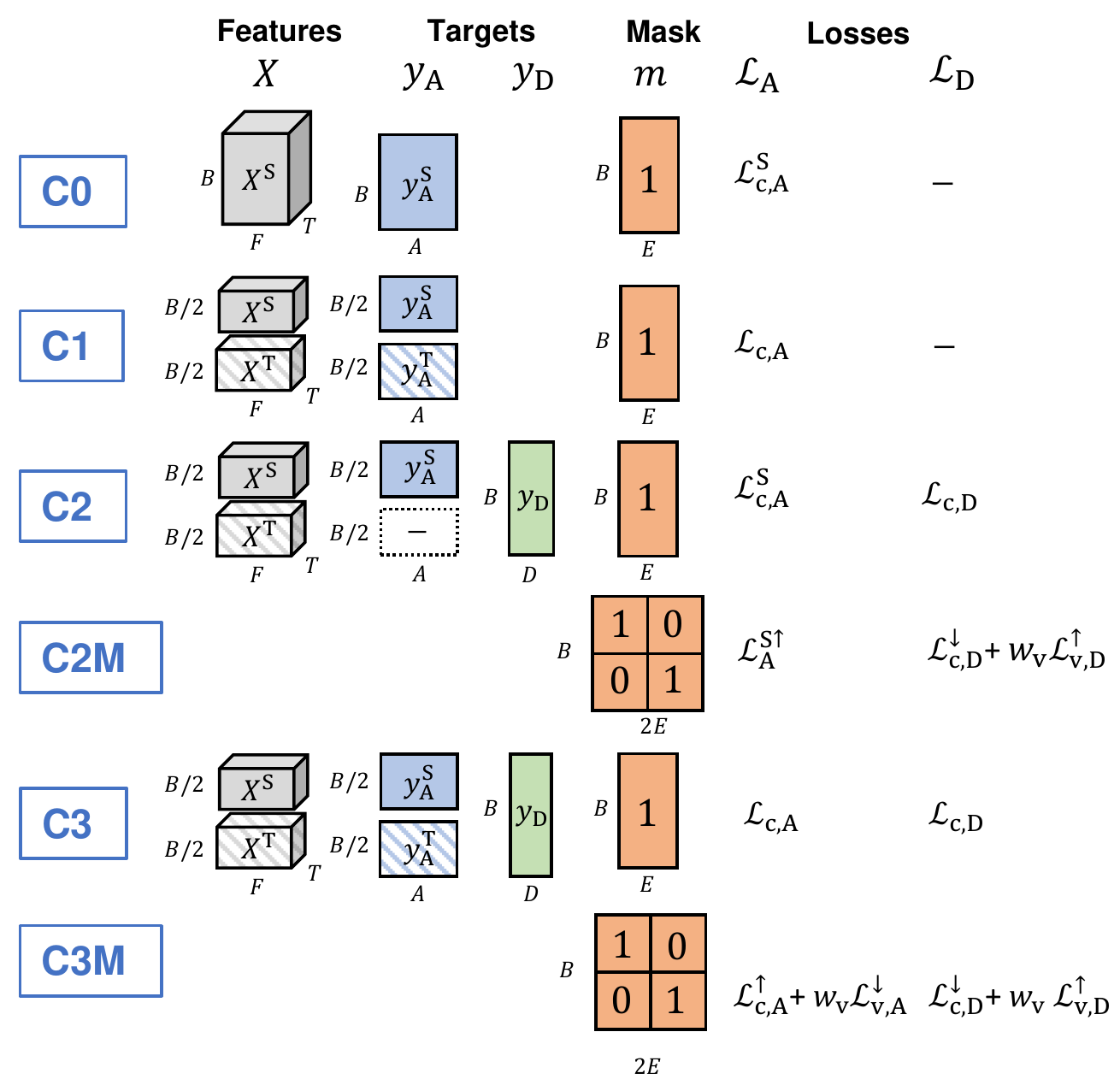}
		\caption{Overview of different loss configurations. For each configuration, the composition of features, targets, and masks per training batch as well as the composition of the two losses $\mathcal{L}_A$ and $\mathcal{L}_D$ is shown (see  \secref{sec:training_loss_configurations} for more details). For the masks (orange), ``1'' and ``0'' indicate all-ones and all-zeroes blocks, respectively.}
		\label{fig:configs}
	\end{center}
\end{figure}

In our work, we compare different approaches to train the neural network for ASC alone or jointly for ASC and domain classification (DC).
In particular, we investigate six training strategies as shown in \figref{fig:configs}.
First, \texttt{C0} involves a conventional ASC model training solely on the source domain data without considering the domain classification task (comparable to \cite{Mezza:2021:DomainAdaptation:EUSIPCO}).
In \texttt{C1}, while following the same training objective, we use both source and target domain data in equal share.
In \texttt{C2}, we consider a hybrid form between unsupervised and supervised DA, where ASC annotations are used only for source but not for target domain data (denoted ``U$^*$'' in \tabref{tab:results}).
Therefore, we can only evaluate the ASC task on the source domain data but at the same time can combine both source and target domain data for the additional DC task.
In \texttt{C3}, a supervised DA scenario is simulated with ASC labels available for both source and target domain data.

The embedding masking procedure detailed in \secref{sec:emb_norm_and_masking} is only applied in configurations \texttt{C2M} and \texttt{C3M}.
Here, different parts of the embeddings in a training batch contribute to different loss terms. 
Since in \texttt{C2M} and \texttt{C3M} only the upper half of the embedding vector contributes to the ASC classification, we double in these cases the embedding size (i.\,e., using an embedding size of $2E$ instead of $E$) to remain comparable with the other configurations.
The embedding masks $m$ and the loss functions $\mathcal{L}_\mathrm{A}$ and $\mathcal{L}_\mathrm{D}$ (defined later) are designed in such way that the first half of the embedding vector supports the ASC task but remains agnostic for DC task and vice versa for the second half.

For each configuration, \figref{fig:configs} illustrates in a column-wise fashion the composition of the feature tensor\footnote{The singleton channel dimension is omitted for better readability.} $x$, the acoustic scene target $y_A$, the  domain target $y_\mathrm{D}$, the embedding mask $m$, as well as the calculation of the loss function $\mathcal{L}$. The tensor dimensions are given in terms of the batch size $B$, the number of Mel bands $F$, the number of time frames $T$, and the embedding size $E$.
For each batch, data instances and corresponding targets are randomly sampled from the relevant domains.

The total loss $\mathcal{L}$ is computed as weighted sum of the individual output branch losses $\mathcal{L}_\mathrm{A}$ and $\mathcal{L}_\mathrm{D}$ for both classification tasks: 
\begin{equation}
    \mathcal{L} = \mathcal{L}_\mathrm{A} + w_\mathrm{D}\mathcal{L}_\mathrm{D}.
\label{eqn:total_loss}
\end{equation}
As shown in the final two columns of \figref{fig:configs}, the losses $\mathcal{L}_\mathrm{A}$ and $\mathcal{L}_\mathrm{D}$ are composed of different variations of the categorical cross-entropy loss $\mathcal{L}_\mathrm{c}$ and a variance-based loss $\mathcal{L}_\mathrm{v}$ defined in \eqref{eqn:lc} and \eqref{eqn:lv}.
Empirically, we found the two weighting factors $w_\mathrm{D}=10$ and $w_\mathrm{v}=500$ (compare \texttt{C2M} and \texttt{C3M}, \figref{fig:configs}) to balance out the contribution of the individual loss terms in the weighted sums.

To define these losses, we consider a multi-class classification with $N$ classes, a batch size of $B$, batch-wise network predictions $\hat{y}~\in~\mathbb{R}^{B \times N}$, and the corresponding one-hot encoded targets $y~\in~\mathbb{R}^{B \times N}$.
The categorical cross-entropy loss can be expressed as
\begin{equation}
    \mathcal{L}_\mathrm{c}(\hat{y}, y) = - \frac{1}{B} \sum_{b=1}^B\sum_{n=1}^N y_{b,n} \log \hat{y}_{b,n}.
\label{eqn:lc}
\end{equation}
This loss function guides the optimization of a neural network to improve its performance for a given classification task.
Opposed to that, the variance-based loss 
\begin{equation}
    \mathcal{L}_\mathrm{v}(\hat{y}) = - \frac{1}{B} \sum_{b=1}^B \frac{1}{N}\sum_{n=1}^N (\hat{y}_{b,n} - \overline{\hat{\mu}_{b}})^2. 
\label{eqn:lv}
\end{equation}
when applied on the networks softmax predictions maximizes the uncertainty for a given classification task. The average of all class predictions is denoted as $\overline{\hat{\mu}_{b}}$. Losses which are only computed using the first and second half of each mini batch are indicated by the superscripts $\uparrow$ and $\downarrow$, respectively.

\section{Evaluation}

 \begin{table*}[t]
 \centering
\scalebox{1}{  \begin{tabular}{p{.1\linewidth}p{.04\linewidth}p{.06\linewidth}p{.06\linewidth}p{.06\linewidth}p{.06\linewidth}p{.06\linewidth}p{.06\linewidth}p{.06\linewidth}p{.06\linewidth}}
    \toprule
    \textbf{Config} & \textbf{DA} & \multicolumn{4}{l}{\textbf{ASC}} & \multicolumn{4}{l}{\textbf{DC}}  \\
      & 
      & \multicolumn{2}{c}{\textbf{$a^{\mathrm{S}}_\mathrm{A}$}} &
      \multicolumn{2}{c}{\textbf{$a^{\mathrm{T}}_\mathrm{A}$}} &
      \multicolumn{2}{c}{\textbf{$a^{\mathrm{S}}_\mathrm{D}$}} &
      \multicolumn{2}{c}{\textbf{$a^{\mathrm{T}}_\mathrm{D}$}}  \\
      & & \textbf{($a^{\mathrm{S}+}_\mathrm{A}$)} & \textbf{($a^{\mathrm{S}-}_\mathrm{A}$)} & \textbf{($a^{\mathrm{T}+}_\mathrm{A}$)} & \textbf{($a^{\mathrm{T}-}_\mathrm{A}$)} & \textbf{($a^{\mathrm{S}+}_\mathrm{D}$)} & \textbf{($a^{\mathrm{S}-}_\mathrm{D}$)} & \textbf{($a^{\mathrm{T}+}_\mathrm{D}$)} & \textbf{($a^{\mathrm{T}-}_\mathrm{D}$)} \\
       
       \midrule
    \multicolumn{10}{l}{\textbf{Related Work}} \\
    \midrule
    \cite{Gharib:2018:DomainAdaptationASC:DCASE} & U$^*$ & \multicolumn{2}{c}{0.65} & \multicolumn{2}{c}{0.32} & \multicolumn{2}{c}{-} & \multicolumn{2}{c}{-}\\
    \cite{Drossos:2019:DomainAdaptation:WASPAA} & U$^*$ & \multicolumn{2}{c}{0.64} & \multicolumn{2}{c}{0.45} & \multicolumn{2}{c}{-}& \multicolumn{2}{c}{-}\\
    \cite{Mezza:2021:DomainAdaptation:EUSIPCO} & U & \multicolumn{2}{c}{0.66} & \multicolumn{2}{c}{\textbf{0.51}} & \multicolumn{2}{c}{-} & \multicolumn{2}{c}{-}\\
    \midrule
    \multicolumn{10}{l}{\textbf{Proposed Method}} \\
    \midrule

    \texttt{C0} & - & \multicolumn{2}{c}{0.66} & \multicolumn{2}{c}{0.19} & \multicolumn{2}{c}{-} & \multicolumn{2}{c}{-} \\

    \texttt{C0}$^{[12]}$ & U & \multicolumn{2}{c}{0.66} & \multicolumn{2}{c}{\textbf{0.46}} & \multicolumn{2}{c}{-} & \multicolumn{2}{c}{-} \\

    \texttt{C1} & S & \multicolumn{2}{c}{0.64} & \multicolumn{2}{c}{0.45} &\multicolumn{2}{c}{-} & \multicolumn{2}{c}{-} \\

    \texttt{C2} & U$^*$ & \multicolumn{2}{c}{0.61} & \multicolumn{2}{c}{0.40} & \multicolumn{2}{c}{1.00} & \multicolumn{2}{c}{0.02} \\

    \texttt{C2M} & U$^*$ & 0.63 & 0.08 & 0.38 & 0.09 & 0.29 & 1 & 0.41 & 0.03 \\

    \texttt{C3} & S &  \multicolumn{2}{c}{0.62} & \multicolumn{2}{c}{0.37} & \multicolumn{2}{c}{1.00} & \multicolumn{2}{c}{0.02} \\

    \texttt{C3M} & S &  0.62 & 0.12 & 0.39 & 0.12 & 0.71 & 1 & 0.16 & 0.01 \\

    \bottomrule
  \end{tabular}}
  \label{tab:results}
 \caption{Experimental results for the ASC (A) and DC (D) tasks for the loss configurations described in \secref{sec:training_loss_configurations}. Accuracy values are provided for both source (S) and target (T) domain data taken from the test set. Results of prior work on unsupervised DA on the same dataset from \cite{Gharib:2018:DomainAdaptationASC:DCASE, Drossos:2019:DomainAdaptation:WASPAA, Mezza:2021:DomainAdaptation:EUSIPCO} are added as reference. The second column indicates whether the DA is unsupervised (U), supervised (S), or a hybrid form of both (U$^*$) (see \secref{sec:training_loss_configurations}) For both configurations \texttt{C2M} and \texttt{C3M}, which include the embedding masking step (see \secref{sec:emb_norm_and_masking}), accuracy values are provided both embedding mask settings (additional superscript ``+'' indicates that $m$ is filled with ones in its first half and with zeros in its second half vice versa for ``-'') to illustrate the task disentanglement. 
 }
 \end{table*}
 
In our experiments, we use an embedding size of $E=256$ and a batch size of $B=512$.
The Adam optimizer is used for training with an initial learning rate of $10^-3$ for 400 epochs. The learning rate is halved every 50 epochs.  
In order to better evaluate the effectiveness of the proposed disentanglement approaches, we avoid any data augmentation and model regularization during training.
We evaluate six models trained with the loss configurations introduced in \secref{sec:training_loss_configurations}.
Furthermore, for the configuration \texttt{C0}, which does not involve any adaptation to the target domain, we investigate the effectiveness of the unsupervised band-wise adaptation proposed by Mezza et al.~\cite{Mezza:2021:DomainAdaptation:EUSIPCO} (entitled \texttt{C0}$^{[12]}$). Here, the target domain data in the test set is adapted to the source domain data in the training set.

In \tabref{tab:results}, the experimental results are reported as classification accuracy values for the ASC (A) and DC (D) tasks. These values are shown individually for the source domain (S) and target domain (T) parts of the test set.
The ASC accuracy values of the related work \cite{Gharib:2018:DomainAdaptationASC:DCASE, Drossos:2019:DomainAdaptation:WASPAA, Mezza:2021:DomainAdaptation:EUSIPCO} are provided as baseline systems.

We make the following observations.
First, the effectiveness of the domain adaptation proposed by Mezza et al.~\cite{Mezza:2021:DomainAdaptation:EUSIPCO} has been verified, as it increases the ASC target domain accuracy $a_\mathrm{A}^\mathrm{T}$ from 0.19 (without any domain adaptation) to 0.46, which is only slightly below the accuracy value of 0.51 reported in \cite{Mezza:2021:DomainAdaptation:EUSIPCO}. 
Second, the results from configuration \texttt{C1} show that by simply combining training data from both domains, one cannot outperform this unsupervised domain adaptation approach, which does not require any target domain training data.
Third, throughout all configurations \texttt{C2}, \texttt{C2M}, \texttt{C3}, and \texttt{C3M}, the model makes perfect DC predictions for the source domain test data while failing to generalize to the target domain. 

In order to get a better insight into the effectiveness of the proposed embedding masking (see \secref{sec:emb_norm_and_masking}) for task disentanglement, we report individual accuracy values for all task-domain pairs for both embedding mask settings in \texttt{C2M} and \texttt{C3M}. An additional superscript ``+'' indicates that the embedding vector $m$ is filled with ones in its first half and zeros in the second half vice versa for ``-''. 
We observe that both the ASC and DC tasks were disentangled to a certain degree since the upper embedding half shows clearly higher accuracy values for the ASC task (compare $a^{\mathrm{T}+}_\mathrm{A}$ against $a^{\mathrm{T}-}_\mathrm{A}$ and $a^{\mathrm{S}+}_\mathrm{A}$ against $a^{\mathrm{S}-}_\mathrm{A}$) while the lower embedding half shows a better performance on the DC task (compare $a^{\mathrm{S}+}_\mathrm{D}$ against $a^{\mathrm{S}-}_\mathrm{D}$). 
However, despite the task disentanglement, both models do not generalize at all to the target domain for the DC task as it can be seen for $a^{\mathrm{T}+}_\mathrm{D}$.
Finally, adding the embedding masking shows a minor performance improvement on the target domain test set for the supervised DA scenario (compare $a^{\mathrm{T}+}_\mathrm{A}$ in \texttt{C3M} against $a^{\mathrm{T}}_\mathrm{A}$ in \texttt{C3}), while the opposite can be observed for the unsupervised DA scenario (\texttt{C2M} against \texttt{C2}).

\section{Conclusion}

In this paper, we adapt and investigate a disentanglement learning approach based on embedding masking for the task of acoustic scene classification.
We propose to combine this method with a combination of cross-entropy and variance based losses in order to better disentangle the microphone characteristics (audio domain) and the acoustic scene.
To get a better insight into the effectiveness of the disentanglement learning approach, we conduct a systematic study on six different training configurations, which model different unsupervised and supervised DA scenarios.
Our results show that while the two tasks were disentangled in the internal embedding representations, the generalization towards target domain data could only slightly be improved in a supervised DA setting.
Furthermore, we could confirm the effectiveness of a state-of-the-art unsupervised domain adaptation approach \cite{Mezza:2021:DomainAdaptation:EUSIPCO}, which---in comparison to the proposed method---performs an across-domain adaptation of the data in the feature space.
A possible research question for future work is how to best combine such adaptation strategies both at the feature level as well as the level of internal data representations within the ASC model.

\section*{Acknowledgements}

This research was partially supported by H2020 EU project AI4Media---A European Excellence Centre for Media, Society and Democracy---under Grand Agreement 95191 and by the Fraunhofer  Innovation Program ``SEC Learn FLY''. The authors would like to express gratitude to Alessandro Ilic Mezza, Sebastian Ribecky, and Sascha Grollmisch for valuable discussions.
The International Audio Laboratories Erlangen are a joint institution of the Friedrich-Alexander-Universit\"at Erlangen-N\"urnberg (FAU) and the Fraunhofer Institute for Integrated Circuits IIS.

\bibliographystyle{IEEEbib}
\bibliography{refs_abr}

\end{document}